\documentclass[a4paper,prb,showpacs,superscriptaddress]{revtex4}
\usepackage{amsfonts}
\usepackage{amssymb}
\usepackage{amsmath}
\usepackage{graphicx}
\def\zk{\bf}
\usepackage{psfig}
\usepackage{bm}

\begin{document}

\def\Green{} 
\def\Black{} 

\def\zk#1{\Green#1\Black}  

\title{Entanglement in a two-qubit Ising model under a site-dependent
external magnetic field}

\author{A. F. Terzis \footnote{E-mail: terzis@physics.upatras.gr}}

\affiliation{Physics Department, School of Natural Sciences,
University of Patras, Patras 265 04, Greece}

\author{E. Paspalakis \footnote{E-mail: paspalak@upatras.gr}}

\affiliation{Materials Science Department, School of Natural
Sciences, University of Patras, Patras 265 04, Greece}

\date{\today}

\begin{abstract}
We investigate the ground state and the thermal entanglement in
the two-qubit Ising model interacting with a site-dependent
magnetic field. The degree of entanglement is measured by
calculating the concurrence. For zero temperature and for certain
direction of the applied magnetic field, the quantum phase
transition observed under a uniform external magnetic field
disappears once a very small non-uniformity is introduced.
Furthermore, we have shown analytically and confirmed numerically
that once the direction of one of the magnetic field is along the
Ising axis then no entangled states can be produced,
independently of the degree of non-uniformity of the magnetic
fields on each site.
\end{abstract}


\maketitle

\section{INTRODUCTION}

The existence of entangled states of component systems and their unique
properties have attracted a lot of attention since the early days of quantum
mechanics [1-5]. Entanglement has also recently been the subject of several
investigations as it plays an important role in the topical areas of quantum
computation and quantum information processing [6,7].

The Heisenberg magnetic spin system is one of the few physical systems that
entanglement arise naturally. Nielsen [7] was the first to report results on
entangled states utilized in a two-spin system. He calculated the
entanglement of formation of a magnetic system described by an isotropic XXX
Heisenberg Hamiltonian in an external magnetic field directed along the
z-axis. Later, Arnesen \textit{et al.} [8] systematically investigated the dependence of
entanglement on temperature and on the applied external field in the same 1D
Heisenberg system. They showed that for ferromagnets the spins are always
disentangled while entanglement is observed for antiferromagnets. In
addition, they found that the degree of entanglement can be enhanced by
increasing the external parameters (magnetic field and/or temperature).
Then, Wang [9,10] studied the effect of anisotropy on thermal entanglement,
working on the two-qubit quantum anisotropic Heisenberg XY model [9] and on
the anisotropic Heisenberg XXZ model [10]. He also investigated the
isotropic Heisenberg XX model with an external magnetic field applied along
the $z$-axis [9]. Kamta and Starace [11] investigated the thermal entanglement
of a two-qubit Heisenberg XY chain in the presence of an external magnetic
field along the $z$-axis. They showed that by adjusting the magnetic field
strength, entangled states are produced for any finite temperature. Sun \textit{et al.}
[12] extended later the work reported in Ref. [11] by introducing a
non-uniform magnetic field. Comparing to the uniform field case, they showed
that entanglement can be more effectively controlled via a non-uniform
magnetic field. The full anisotropic XYZ Heisenberg spin two-qubit system in
which a magnetic field is applied along the $z$-axis, was studied by Zhou \textit{et al.}
[13]. The enhancement of the entanglement for particular fixed magnetic
field by increasing the $z$-component of the coupling coefficient between the
neighboring spins, was their main finding. Finally, thermal entanglement in
a two-qubit Ising model assuming an applied magnetic field in an arbitrary
direction has been investigated by Gunlycke \textit{et al.} [14].

Several, mainly numerical, works exist also on the study of
pairwise thermal entanglement in the $n$-qubit Heisenberg spin
chain. In short, the systems being studied are the XXZ
three-qubit Heisenberg model with an applied magnetic field in
the $z-$ direction [15], the XYZ three-qubit Heisenberg model
with an applied magnetic field along the $z$-axis [13], the XX
four-qubit Heisenberg model with an applied magnetic field in the
$z$-direction [16], the XX [17] and XXX and XXZ [18] four- and
five-qubit Heisenberg model and the XX $m$-qubit (with $m$ up to
eleven) Heisenberg model [19].

It turns to be quite interesting to study the general case of
different magnetic fields at each spin site. The control of the
applied field at each spin separately is very useful in order to
perform quantum computations [20]. Hence in the present
theoretical analysis we investigate the ground state and the
thermal entanglement in an Ising model with a non-uniform and
anisotropic external magnetic field, i.e. the magnitude and the
direction of each magnetic field is different in each spin.

This article is organized as follows: In the following section we
present the details of the theoretical analysis based on the
calculation of the concurrence [21,22] of the system. Then, in
section III numerical results are presented and discussed for
several cases of the parameters of the system (magnetic fields
magnitudes and directions and temperature). In addition, a
general analytical result is presented in the Appendix. There, we
show that once one of the applied external magnetic fields points
along the Ising direction, no matter what is the direction and
magnitude of the other field the concurrence is always zero. This
result is valid for zero and finite temperatures. Our results are
summarized in section IV.

\section{THEORY}

The Hamiltonian studied in this work is given by
\begin{equation}
H = J(\sigma _1^z \sigma _2^z + \sigma _2^z \sigma _1^z ) + B_1
\cos \theta _1 \sigma _1^z + B_1 \sin \theta _1 \sigma _1^x + B_2
\cos \theta _2 \sigma _2^z + B_2 \sin \theta _2 \sigma _2^x \quad
,
\end{equation}
where $\sigma^a$ are the Pauli operators and $J$ is the strength
of the Ising interaction. Also, $B_{1}$ and $B_{2}$ are the
magnitudes of the external magnetic fields. We assume that each
magnetic field has an arbitrary direction, defined by the angles
$\theta _1 $ and $\theta _2 $ between the field and the Ising
direction. It is sufficient to consider that the magnetic field
lies in a plane (\textit{xz}) containing the Ising $z$-direction,
because in three spatial dimensions, the Hamiltonian possesses
rotational symmetry about the $z$-axis.

A useful and convenient quantitative tool, which has been developed to study
entanglement, is the entanglement of formation [4]. The entanglement of
formation of a state of a composite system is proportional to the minimum
number of Bell states, which must be shared between the components of the
composite system. There is no general prescription for evaluation of the
entanglement of formation in arbitrary systems. In this area Wootters [21]
has described an explicit method for evaluating the entanglement of
formation of an arbitrary state of a two-qubit system. His explicit formula
is a generalization of the expression derived by him and Hill [22] for a
special class of density matrices.

For a pair of qubits the entanglement of formation $E_{12}$ is
estimated from the expression $E_{12} = h\left( {1 + \sqrt {1 -
C_{12}^2 } } \right) / 2$, where $C_{12}$ denotes the concurrence
and $h$ is the binary entropy function [21]. The concurrence is
defined as $C_{12} = \max \{\lambda _1 - \lambda _2 - \lambda _3
- \lambda _4 ,0\}$, where the $\lambda $'s are the square roots
of the eigenvalues in decreasing order of magnitude (i.e.
$\lambda _1 \ge \lambda _2 \ge \lambda _3 \ge \lambda _4 )$ of
the spin-flipped density matrix operator $R_{12} = \rho (\sigma
^y \otimes \sigma ^y)\rho \ast (\sigma ^y \otimes \sigma ^y)$.
Here, $\rho $ is the density matrix operator defined as $\rho =
\exp ( - H / k_B T) / Z$, where $Z$ is the partition function ($Z
= tr\{\exp ( - H / k_B T)\})$ and $k_B $ is the Boltzmann's
factor. As concurrence is a monotonically increasing function of
$E_{12} $ and both functions have values in the range 0 to 1, we
practically use $C_{12} $ as a measure of the entanglement. Zero
concurrence corresponds to an un-entangled pair of states and
unity concurrence to a maximally entangled pair of states. This
type of entanglement is usually called thermal entanglement as it
is described by a temperature dependent density matrix operator.

In order to proceed we need to find the eigenvalues ($E_{i})$ and
eigenvectors ($\left| {\Psi _i } \right\rangle )$ of the
Hamiltonian of the Ising system in the presence of a non-uniform
external magnetic field, i.e. the Hamiltonian of Eq. (1). Once we
have determined the eigenstates of the system, the density matrix
operator can be written as
\begin{equation}\rho = Z^{ -
1}\sum\limits_i {\exp ( - E_i / k_B T)} \left| {\Psi _i }
\right\rangle \left\langle {\Psi _i } \right| \quad .
\end{equation}
Then, the spin-flipped matrix operator is evaluated in a $4 \times
4$ matrix representation, in terms of the natural basis vectors
$\{\left| {00} \right\rangle ,\left| {01} \right\rangle ,\left|
{10} \right\rangle ,\left| {11} \right\rangle \}$. In most cases,
even if one obtains analytic expressions for the eigenvalues of
the spin-flipped matrix operator, it is practically impossible to
derive a simple analytic expression for the concurrence. The
reason lies to the fact that the relative order of magnitude of
the eigenvalues of $R_{12,}$ depends on the parameters involved.
In general the concurrence can be evaluated numerically. For
particular values of the parameters of the system, an analytic
expression can be achieved (see in the Appendix). It is worth
pointing out that the above analysis is greatly simplified for
the zero temperature case, and then only the ground state is
populated and hence $\rho = \left| {\Psi _0 } \right\rangle
\left\langle {\Psi _0 } \right|$, where 0 is the index that
denotes the ground state of the system.

\section{RESULTS AND DISCUSSION}

We begin our discussion with the results of the ground state of
the system at zero absolute temperature $\left( {T = 0} \right)$.
In Fig. 1(a) we assume a uniform magnetic field (i.e. same
magnitude and same direction of magnetic field in each spin) and
present the results of the concurrence as a function of the
strength of the applied fields. It is clear that the entanglement
is highest for nearly vanishing magnetic fields and decreases
with increasing the field's amplitude. In this figure we observe
that for $B_1 = B_2 = B \to 0^ + $, the concurrence is unity,
which shows the creation of maximally entangled states. Once we
set a zero value for the magnetic field, the eigenstates are the
same as those of the Ising model without magnetic field, i.e. the
standard disentangled basis $\{\left| {00} \right\rangle ,\left|
{01} \right\rangle ,\left| {10} \right\rangle ,\left| {11}
\right\rangle \}$, as the Hamiltonian is diagonal in this basis.
Therefore, the concurrence obtains a zero value. Then, even for an
infinitesimal increase of the magnetic field, the system goes
from a non-entangled state to a maximally entangled state. This
is a clear evidence of a quantum phase transition (QPT) [23]. In
the case that the magnetic field is along the $z$-axis, so that
$\theta _1 = \theta _2 = 0$, we have no entanglement at all. In
this case the ground state has energy --2$J$ and is doubly
degenerate. For small, but equal, angles $\theta _1 = \theta _2 $,
there are two energy levels, one of energy value --2$J$ and the
other with energy close to --2$J$, with corresponding states, the
first one equal to the Bell state $\left( {\left| {01}
\right\rangle - \left| {10} \right\rangle } \right) / \sqrt 2 $
and the other close to the Bell state $\left( {\left| {01}
\right\rangle + \left| {10} \right\rangle } \right) / \sqrt 2 $.
Thus, we get a maximally entangled qubit pair. Hence, we conclude
that even a very small component of the magnetic field along the
$x$-axis is adequate to create entangled states. In all cases we
observe that the concurrence drops to zero for very strong
fields. This is expected as for very strong fields the spins will
be completely aligned along the field direction and hence the
entanglement will drop to zero. It is worth noting that the
smaller the $x$-component of the magnetic field becomes, the
faster the concurrence drops to zero for $B>2J$. In the case
where the magnetic field is along the $x$-axis the concurrence
can be calculated analytically from the density matrix of the
pure (ground) state, and we get $C_{12} = \left[ {1 + \left( {B
\mathord{\left/ {\vphantom {B J}} \right.
\kern-\nulldelimiterspace} J} \right)^2} \right]^{ - 1 / 2}$, see
short dashed line ($\theta _1 = \theta _2 = \pi / 2)$ in Fig.
1(a). The physical explanation of this behavior comes from the
fact that the field tends to align the qubit spins in a different
disentangled state from the spin-spin coupling. This implies that
it is the trade off between the field and the Ising interaction
that produces the entanglement.

We now investigate the case of non-uniform magnetic field, which is actually
more realistic than the uniform magnetic field case. In Fig. 1(b) we observe
that for $B_2 $ slightly different from $B_1 $ (here, $B_2 = 1.0005[B_1
\equiv B])$ the QPT disappears in cases of small $x$-component of the magnetic
fields. Now the concurrence starts from vanishing value reach a maximum
value for $B<2J$ and then drops to zero very abruptly for $B>2J$. For small, but equal,
angles $\theta _1 = \theta _2 $, there are two energy levels one of energy
value --2$J$ and the other with energy value close to --2$J $but in this case the
corresponding states are not close to the Bell states $\left( {\left| {01}
\right\rangle \pm \left| {10} \right\rangle } \right) / \sqrt 2 $. As the
$x$-component of the magnetic field increases the change from the vanishing
value of the concurrence at zero field is very abrupt, as can be seen from
the short dashed curve ($\theta _1 = \theta _2 = \pi / 2)$ in Fig. 1(b). For
example, for $\theta _1 = \theta _2 = 0.3\pi $ and for a magnetic field of
value $B=$0.01$J$ the concurrence is practically unity. A very similar behavior is
observed for slightly different directions of the magnetic fields $B_1 $and
$B_2 $, as can be seen in Fig.1(c). In this case, we observe that the QPT
disappears for magnetic fields of the same magnitude but slightly different
direction. It is important to point out that this phenomenon is present even
for directions of the magnetic fields very close to the $x$-direction. Again,
we get that there are two lowest energy levels corresponding to states that
are not close to the Bell states $\left( {\left| {01} \right\rangle \pm
\left| {10} \right\rangle } \right) / \sqrt 2 $. For larger differences
between the magnitudes of the magnetic fields for the two qubits, i.e. $B_2
= 1.05B_1 $, the disappearance of the QPT is even more pronounced for cases
characterized by small $x$-component of the magnetic fields [see solid curve in
Fig. 1(d)]. A systematic study of the field effects is depicted in Fig.2,
where we plot the concurrence for cases of small [Figs. 2(a) and (b)] and
large [Figs. 2(c) and (d)] difference in the magnetic field magnitude. We
observe that the entanglement achieved for low $B_x $ components is the one
that it is more influenced. For example, even for a very small difference in
the field amplitudes (of the order of 1.05) the maximum entanglement is
diminished. For a larger value of the amplitudes of the field ratio, e.g.
for 1.5, the concurrence practically disappears (not shown). The reason for
this behavior is due to the fact that the eigenvectors are in this case the
non-entangled pure states $\left| {01} \right\rangle ,\left| {10}
\right\rangle $ and $\left| {00} \right\rangle $. As we see from the same
figure, the influence of the non-uniformity of the magnetic fields becomes
smaller the more the direction of the fields get closer to a direction along
the $x$-axis. In the same figure we investigate the influence of the difference
in the field directions. We note that the maximum change in the behavior is
obtained for the case that $B_x $ is small.

We will now study the case of finite temperature, i.e. the case of thermal
entanglement. We examine how the effects discussed in Fig. 1(a) changes for
two different values of temperature with $B_1 = B_2 ( = B)$, $\theta _1 =
\theta _2 $. From Figs. 3 and 4 we clearly see that the critical temperature
depends on the orientation of the fields and that the field at which the
maximum entanglement occurs as the temperature increases shifts to larger
values of the magnetic field amplitudes. First, we study the case of the low
$B_x $ component, where we observe a sharp peak in the concurrence. We note
there is a fast drop of the concurrence to zero for a low $B_x $ component
and for low values of the field magnitude. The behavior found in Figs 3 and
4 has already reported and discussed in Ref. [14]. From the same plots we
observe that as the direction of the fields gets close to an orthogonal
direction with respect to the Ising axis, the maximum entanglement occurs at
lower values of the magnetic filed magnitude. In the case of $\theta = \pi /
2$, the concurrence does not drop to zero very fast as the eigenvectors of
all eigenstates are either Bell states or a linear combination of Bell
states.

Comparing Fig. 1(a) and Fig. 3(a), we clearly see that the QPT is not
present in the finite temperature case. Moreover, we perform calculations
for finite temperature with magnetic fields the same as those in Figs 1(b)
and 1(c). We have found that for these values of the external parameters,
the concurrence is practically the same as in Fig.3(a). This finding is
valid once the magnitudes of the two applied fields are very similar. For
case of larger difference in the field magnitude, the finite temperature
results are more influenced as one sees by comparing Fig.1(d) and Fig.3(b).
In this case of finite temperature under higher anisotropy of fields the QPT
disappears, independently of the direction of the applied fields. Another,
general feature seen from Fig. 4 (in comparison to Fig. 2) is that in all
the cases the finite temperature leads to a decrease in the concurrence and
shifts the maximum value of the concurrence at stronger applied fields. The
effect is more pronounced for directions of the fields along an axis
perpendicular to the Ising axis.

It is known [14] that there is an angle $\theta _1 = \theta _2 = \theta
\left( {B,T} \right)$ where the entanglement is maximum for a given
temperature and amplitude ($B_1 = B_2 )$. This feature, known as the
phenomenon of magnetically induced entanglement, has been explained
heuristically assuming that with $B_x $ and $B_z $ fixed, the entanglement
should change continuously with temperature. As the increase in the
temperature widens the low-entanglement zone around $B_x = 0$ [14] and the
entanglement has to fall for large $B_x $, it is expected that at some
intermediate value of $B_x $, the maximal entanglement will be reached. The
preferred angle traverses from $\theta = 0$ at zero temperature to $\theta =
\pi / 2$ at $T\sim J / k_B $. First, we investigate the zero temperature
case. For parallel directions of the magnetic fields, the maximum
concurrence is achieved for fields pointing along the $x$-axis (not shown
here). Moreover, we observe that once we fix the direction of one of the
applied fields the direction of the field applied at the other spin depends
on the relative magnitude of the fields. This dependence is minimal once one
of the fields is along the $x$-axis (not shown here). We note that there is a
cutoff value of the magnetic field magnitude applied on the second spin,
above which the relative angle remains practically the same. This cutoff
value depends on the direction of the magnetic field applied at the first of
the spins. We have shown that for non-uniform fields ($B_1 = J,\mbox{ }B_2 =
1.5J)$ the maximum concurrence is obtained for non-parallel directions of
the two applied fields. We observe that the smaller the $B_x $ component of
one of the fields becomes the more the system deviates from the parallel
direction. Actually, we have found that for fields of equal magnitudes and
smaller than $2J$, no matter what is the direction of the field applied at the
first spin, the maximum concurrence is achieved for magnetic field at the
second spin pointing at the same direction as in the first one. This is not
true for fields of equal magnitudes but with values larger than $2J$. The
deviation from parallel fields becomes more pronounced the more the
direction of one of the fields gets closer to the Ising direction. It is
well understood that if both fields get very large values the concurrence is
zero. In our study we found that there is no entanglement even if either one
of the fields is very large. How large should the field be in order to
destroy entanglement depends on the orientations of the fields. A rule of
thumb is that we need larger fields as the directions of the fields gets
closer to a direction perpendicular to the Ising direction.

In Fig. 5(a) we confirm that for fields of equal magnitude ($B_1
= B_2 )$, the maximum concurrence occurs for $\theta _1 = \theta
_2 $. On the contrary, Fig. 5(b) shows that for non-uniform
fields the maximum entanglement occurs for unequal angles
($\theta _1 \ne \theta _2 )$. Actually, as the difference in the
magnitude of the two fields increases, the maximum entanglement
occurs for direction of the larger field along the $x$-axis
independent of the direction of the other field. A very similar
behavior was found for a finite temperature as shown in Fig.6.

\section{CONCLUSIONS}

In summary, in the present work, a systematic investigation was
performed on the entanglement in a two-spin Ising model in a
site-dependent magnetic field. One of the most interesting
results is the finding that the QPT observed at zero temperature
when a uniform magnetic field is applied, disappears with the
introduction of a very small difference in the applied fields.
This difference could be either a difference in magnitude or a
difference in the direction of the magnetic fields. Moreover, we
have found that for parallel fields with direction close to the
$z$-axis (Ising direction), small differences in the magnetic
field magnitudes result in very weak entanglement. On the
contrary, a very large asymmetry in the amplitudes of the applied
fields has small effects on the well-entangled states obtained
for fields with large $x$-component. We have also studied the
phenomenon of the magnetically induced entanglement and observe
that for equal magnitudes of the external magnetic fields the
maximum concurrence occurs for parallel directions of the fields.
Once a non-uniformity is introduced the maximally entangled
states obtained for non-parallel orientations of the fields. In
addition, we have shown that the concurrence drops to zero even
if only one of the fields gets very large values. Finally, we
have derived an analytic result valid for ground state
entanglement and thermal entanglement. The analytic result, which
has been confirmed numerically, predicts that we get vanishing
concurrence once the direction of one of the fields is along the
Ising direction.

\section*{APPENDIX}

In the Appendix we discuss the special case of one magnetic field
parallel to $z$-azis and the other in any direction. We find
analytic expressions that predict that the concurrence in this
case is always zero. The Hamiltonian studied in this case is
given by

\begin{equation}
H = J(\sigma _1^z \sigma _2^z + \sigma _2^z \sigma _1^z ) + B_1
\sigma _1^z + B_2 \cos \theta \sigma _2^z + B_2 \sin \theta
\left( {\sigma _1^ + + \sigma _1^ - } \right) / 2 \quad ,
\end{equation}
where $\sigma^a$ are the Pauli operators and $J$ is the strength
of the Ising interaction. Also, $B_{1}$ and $B_{2}$ are the
magnitudes of the external magnetic fields. We assume that the
first spin feels a magnetic field along the $z$-direction and the
second field has an arbitrary direction, defined by the angle
$\theta $ between the field and the Ising direction.

The eigenenergies and the eigenstates can be calculated
analytically. The eigenenergies are:
\begin{eqnarray}
E_\pm ^X &=& - B_1 \pm \left[ {4J^2 + B_2^2 - 4JB_2 \cos \theta }
\right]^{1 / 2} \, , \\
E_\pm ^Y &=& B_1 \pm \left[ {4J^2 + B_2^2 + 4JB_2 \cos \theta }
\right]^{1 / 2} \, ,
\end{eqnarray}
The corresponding normalized eigenvectors are
\begin{eqnarray}
\left| {X_\pm } \right\rangle &=& a_\pm \left| {00} \right\rangle
+ b_\pm \left| {01} \right\rangle \, , \\
\left| {Y_\pm } \right\rangle &=& c_\pm \left| {10} \right\rangle
+ d_\pm \left| {11} \right\rangle \, ,
\end{eqnarray}
where,
\begin{eqnarray}
a_\pm &=& \frac{1}{\sqrt 2 }\frac{B_2 \sin \theta }{\left[ {4J^2 +
B_2^2 - 4JB_2 \cos \theta } \right]^{1 / 4}\left[ {\left( {4J^2 +
B_2^2 - 4JB_2 \cos \theta } \right)^{1 / 2}\pm \left( {B_2 \cos
\theta - 2J} \right)} \right]^{1 / 2}} \, , \\
b_\pm &=& \pm \frac{1}{\sqrt 2 }\frac{\left[ {\left( {4J^2 + B_2^2
- 4JB_2 \cos \theta } \right)^{1 / 2}\pm \left( {B_2 \cos \theta
- 2J} \right)} \right]^{1 / 2}}{\left[ {4J^2 + B_2^2 - 4JB_2 \cos
\theta } \right]^{1 / 4}} \, , \\
c_\pm &=& \frac{1}{\sqrt 2 }\frac{B_2 \sin \theta }{\left[ {4J^2 +
B_2^2 + 4JB_2 \cos \theta } \right]^{1 / 4}\left[ {\left( {4J^2 +
B_2^2 + 4JB_2 \cos \theta } \right)^{1 / 2}\pm \left( {B_2 \cos
\theta + 2J} \right)} \right]^{1 / 2}} \, \\
d_\pm &=& \pm \frac{1}{\sqrt 2 }\frac{\left[ {\left( {4J^2 + B_2^2
+ 4JB_2 \cos \theta } \right)^{1 / 2}\pm \left( {B_2 \cos \theta
+ 2J} \right)} \right]^{1 / 2}}{\left[ {4J^2 + B_2^2 + 4JB_2 \cos
\theta } \right]^{1 / 4}} \, .
\end{eqnarray}

Now the density matrix is estimated by the following expression
\begin{equation}
\rho = Z^{ - 1}[e^{ - \beta E_ - ^X }\left| {X_ - } \right\rangle
\left\langle {X_ - } \right| + e^{ - \beta E_ + ^X }\left| {X_ + }
\right\rangle \left\langle {X_ + } \right| + e^{ - \beta E_ - ^Y
}\left| {Y_ - } \right\rangle \left\langle {Y_ - } \right| + e^{
- \beta E_ + ^Y }\left| {Y_ + } \right\rangle \left\langle {Y_ +
} \right|] \, ,
\end{equation}
where the partition function is defined as
\begin{equation}
Z = e^{ - \beta E_ - ^X } + e^{ - \beta E_ + ^X } + e^{ - \beta
E_ - ^Y } + e^{ - \beta E_ + ^Y } \, .
\end{equation}
and $\beta \equiv 1 / k_B T$. Then, the spin-flipped density
matrix operator $R_{12}$ in the regular basis representation has
the following form
\begin{eqnarray}
R_{12} = \left[ {{\begin{array}{*{20}c}
 A \hfill & C \hfill & 0 \hfill & 0 \hfill \\
 D \hfill & B \hfill & 0 \hfill & 0 \hfill \\
 0 \hfill & 0 \hfill & B \hfill & { - C} \hfill \\
 0 \hfill & 0 \hfill & { - D} \hfill & A \hfill \\
\end{array} }} \right] \, ,
\end{eqnarray}
\noindent where, $A = \rho _{11} \rho _{44} - \rho _{12} \rho
_{34}$, $B = \rho _{22} \rho _{33} - \rho _{12} \rho _{34}$, $C =
\rho _{33} \rho _{12} - \rho _{11} \rho _{34}$ and $D = \rho
_{44} \rho _{12} - \rho _{22} \rho _{34}$. The eigenvalues,
$\lambda $'s of the spin flip density matrix operator are easily
estimated as
\begin{eqnarray}
\lambda _1 &=& \lambda _2 = [(A + B) + \left( {(A - B)^2 + 4CD}
\right)^{1 / 2}] / 2 \quad , \\
\lambda _3 &=& \lambda _4 = [(A + B) - \left( {(A - B)^2 + 4CD}
\right)^{1 / 2}] / 2 \, .
\end{eqnarray}
Hence, the concurrence defined as $C_{12} = \max \{\sqrt {\lambda
_1 } - \sqrt {\lambda _2 } - \sqrt {\lambda _3 } - \sqrt {\lambda
_4 } ,0\}$, is always zero no matter what are the values of $A,
B, C$ and $D$. Therefore, we conclude that independently of the
magnitudes of the magnetic fields and independently of the
direction of one of the fields, if one of the fields points along
the Ising direction there is no entanglement in the system.

It is worth mentioning that the above analysis is greatly
simplified for the zero temperature case as only the ground state
is populated and hence $\rho = \left| {\Psi _0 } \right\rangle
\left\langle {\Psi _0 } \right|$, where $0$ is the index for the
ground state. In this case the only non-vanishing terms in the
density matrix operator are the $\rho _{11} $, $\rho _{22} $ and
$\rho _{12} $. Then, it is rather straightforward to show that
the spin-flip density operator matrix is the zero $4 \times 4$
matrix.
%

\section*{REFERENCES}

[1] E. Schr\"odinger, Proc. Cambridge Phil. Soc., \textbf{31}, 555
(1935).

[2] E. Schr\"odinger, Naturwissenschaften, \textbf{23}, 807
(1935).

[3] J.S. Bell, Physics, \textbf{1}, 195 (1964).

[4] C.H. Benett, D.P. DiVincenzo, J.A. Smolin and W.K. Wooters, Phys. Rev.
A, \textbf{54}, 3824 (1996).

[5] C.H. Benett and D.P. DiVincenzo, Nature (London), \textbf{404}, 247
(2000).

[6] M. A. Nielsen and I.L. Chuang, \textit{Quantum Computation and Quantum Information}, (Cambridge University Press, Cambridge,
2000).

[7] M.A. Nielsen, Ph.D. Thesis, University of New Mexico, 1998; see also
LANL e-print: quant-ph/0011036.

[8] M.C. Arnesen, S. Bose and V. Vedral, Phys. Rev. Lett., \textbf{87},
017901 (2001).

[9] X. Wang, Phys. Rev. A, \textbf{64}, 012313 (2001).

[10] X. Wang, Phys. Lett. A, \textbf{281}, 101 (2001).

[11] G.L. Kamta and A.F. Starace, Phys. Rev. Lett., \textbf{88}, 107901
(2002).

[12] Y. Sun, Y. Chen and H. Chen, Phys. Rev. A, \textbf{68}, 044301 (2003).

[13] L. Zhou, H.S. Song, Y.Q. Guo and C. Li, Phys. Rev. A, \textbf{68},
024301 (2003).

[14] D. Gunlycke, V.M. Kendon, V. Vedral and S. Bose, Phys. Rev. A,
\textbf{64}, 042302 (2001).

[15] X. Wang, H. Fu and A.I. Solomon, J. Phys. A:Math. Gen., \textbf{34},
11307 (2001).

[16] X. Wang, Phys. Rev. A, \textbf{66}, 034302 (2002).

[17] X.-Q. Xi, W.-X. Chen, S.-R. Hao and R.-H. Yue, Phys. Lett. A,
\textbf{300}, 567 (2002).

[18] X. Wang and K. M$\phi $lmer, Eur. Phys. J. D, \textbf{18}, 385 (2002).

[19] X. Wang, Phys. Rev. A, \textbf{66}, 044305 (2002).

[20] Y. Makhlin, G. Schoen and A. Shnirman, Rev. Mod. Phys., \textbf{73},
357 (2001).

[21] W.K. Wootters, Phys. Rev. Lett., \textbf{80}, 2245 (1998).

[22] S. Hill and W.K. Wootters, Phys. Rev. Lett., \textbf{78}, 5022 (1997).

[23] S. Sachdev, \textit{Quantum Phase Transitions}, (Cambridge University Press, Cambridge, 1999).

\pagebreak

\begin{figure}
\begin{center}
\centerline{\hbox{ \psfig{figure=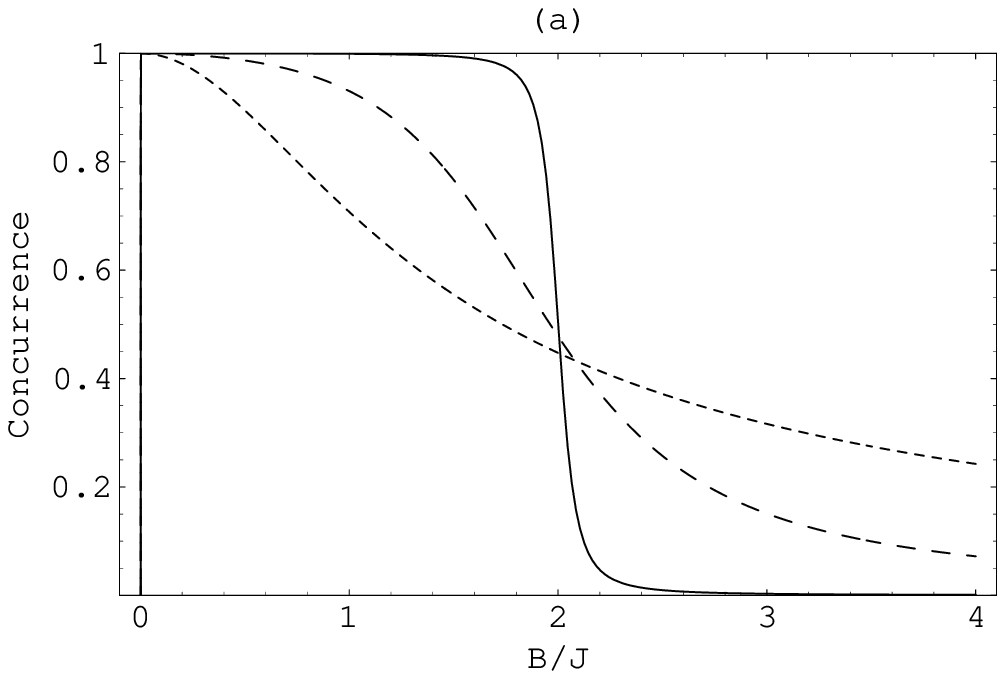,height=5cm}
\psfig{figure=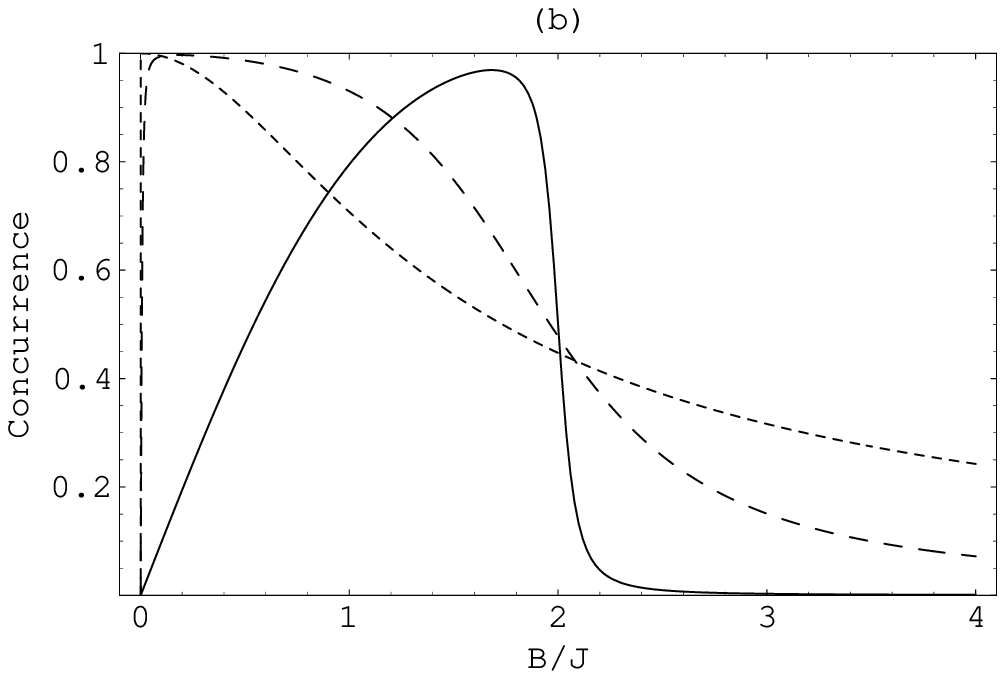,height=5cm}}} \centerline{\hbox{
\psfig{figure=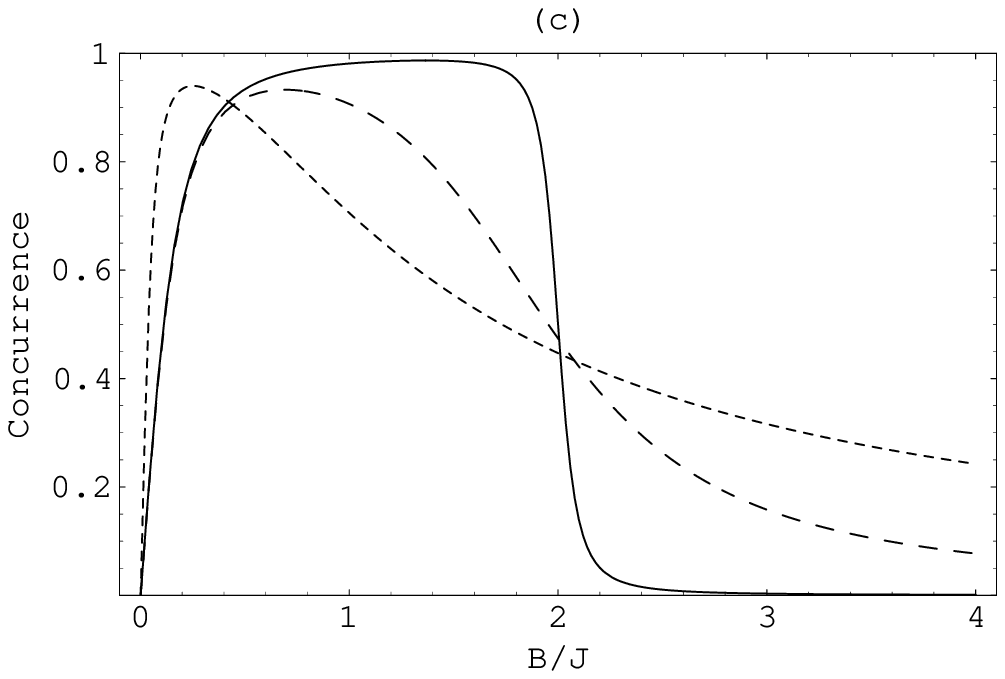,height=5cm}
\psfig{figure=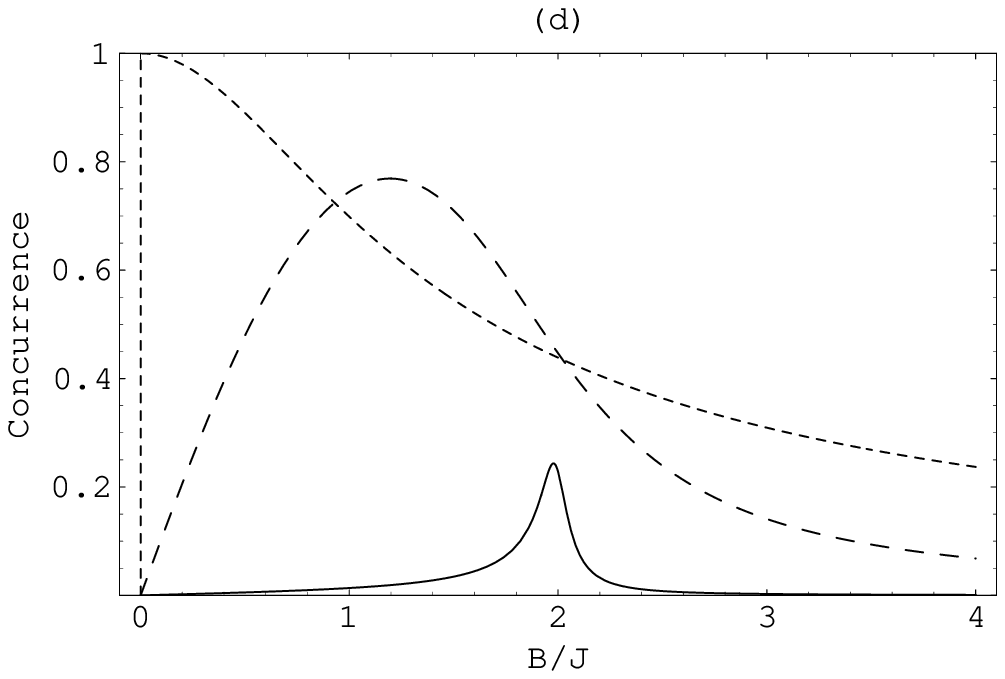,height=5cm}}}
\end{center}
\caption{Plots of concurrence as a function of the magnetic field
magnitude $B$(=$B_{1})$ in units of $J$ at zero temperature.
(\textbf{a}) Equal magnetic field magnitudes ($B_{1}=B_{2}=B)$.
Solid curve is for $\theta _1 = \theta _2 = 0.01\pi $, long and
short dashed curves correspond to $\theta _1 = \theta _2 = 0.1\pi
$ and $\theta _1 = \theta _2 = 0.5\pi $, respectively.
(\textbf{b}) The same as (a) but with unequal field amplitudes
($B_{2}$=1.0005$B_{1})$. (\textbf{c}) Equal magnetic field
magnitudes ($B_{1}=B_{2}=B)$. Solid curve is for $\theta _1 =
0.01\pi ,\theta _2 = 0.011\pi $, long dashed curve is for $\theta
_1 = 0.1\pi ,\theta _2 = 0.11\pi $ and short dashed curve is for
$\theta _1 = 0.5\pi ,\theta _2 = 0.51\pi $. (\textbf{d}) The same
as (a) but with unequal field amplitudes ($B_{2}$=1.05$B_{1})$}
\label{fig1}
\end{figure}

\begin{figure}
\begin{center}
\centerline{\hbox{ \psfig{figure=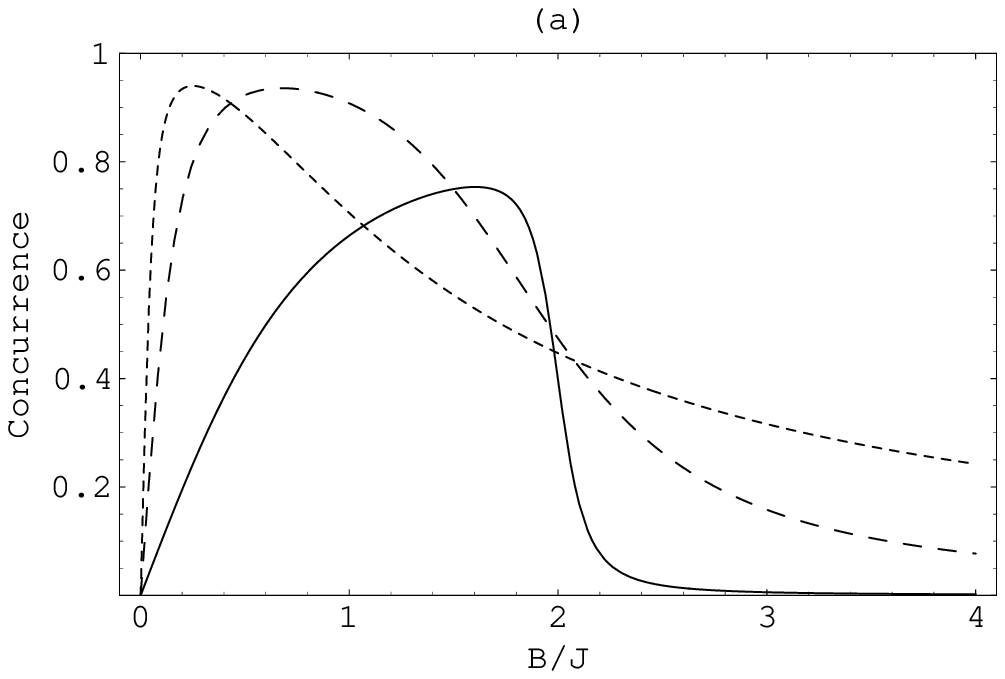,height=5cm}
\psfig{figure=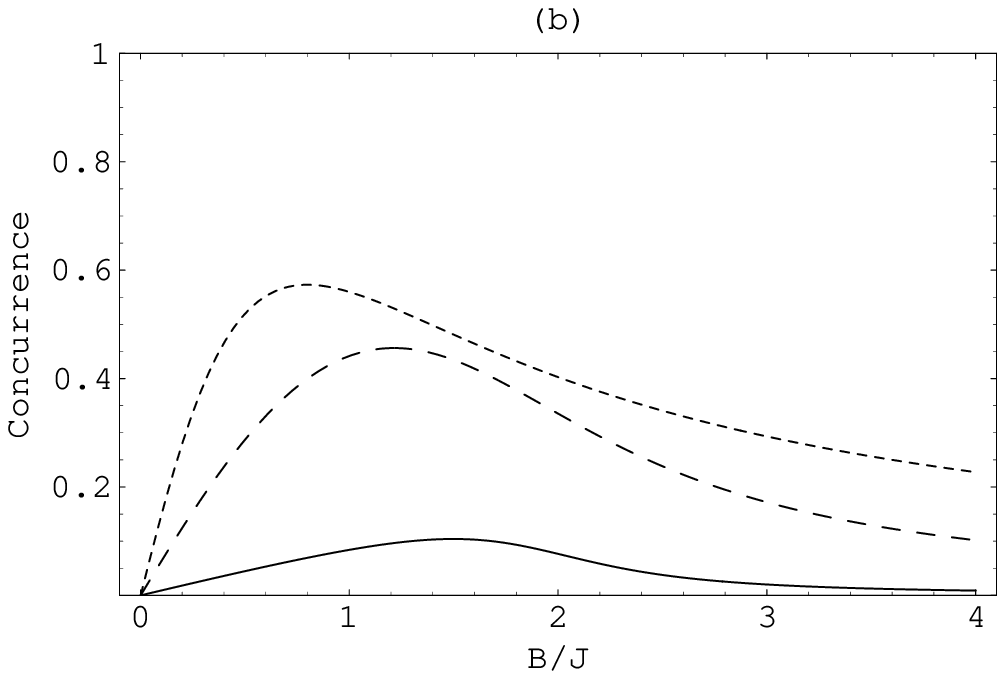,height=5cm}}} \centerline{\hbox{
\psfig{figure=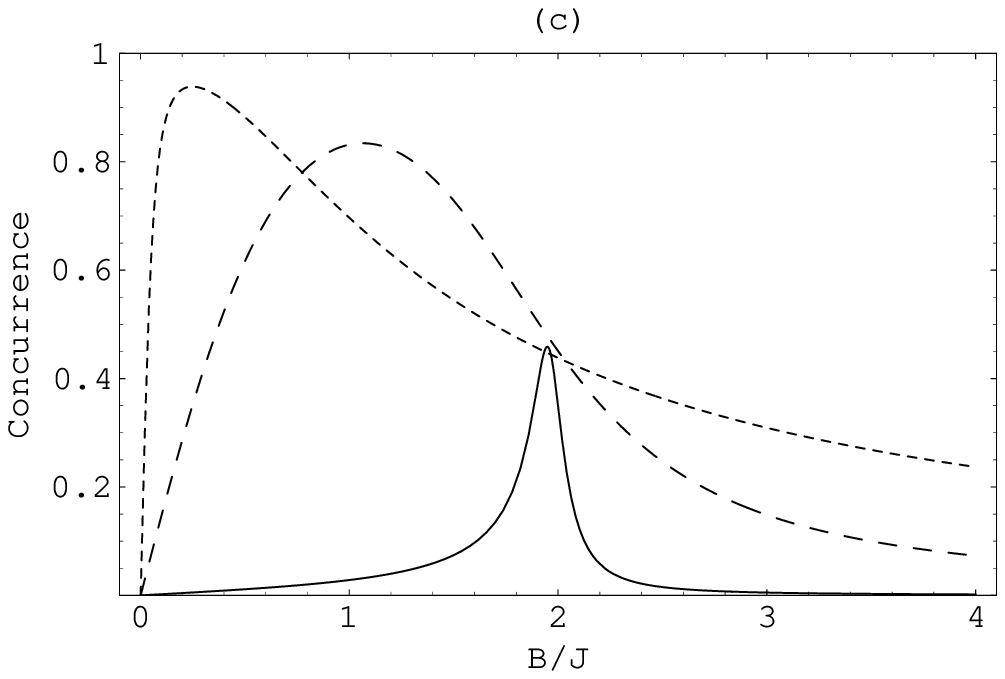,height=5cm}
\psfig{figure=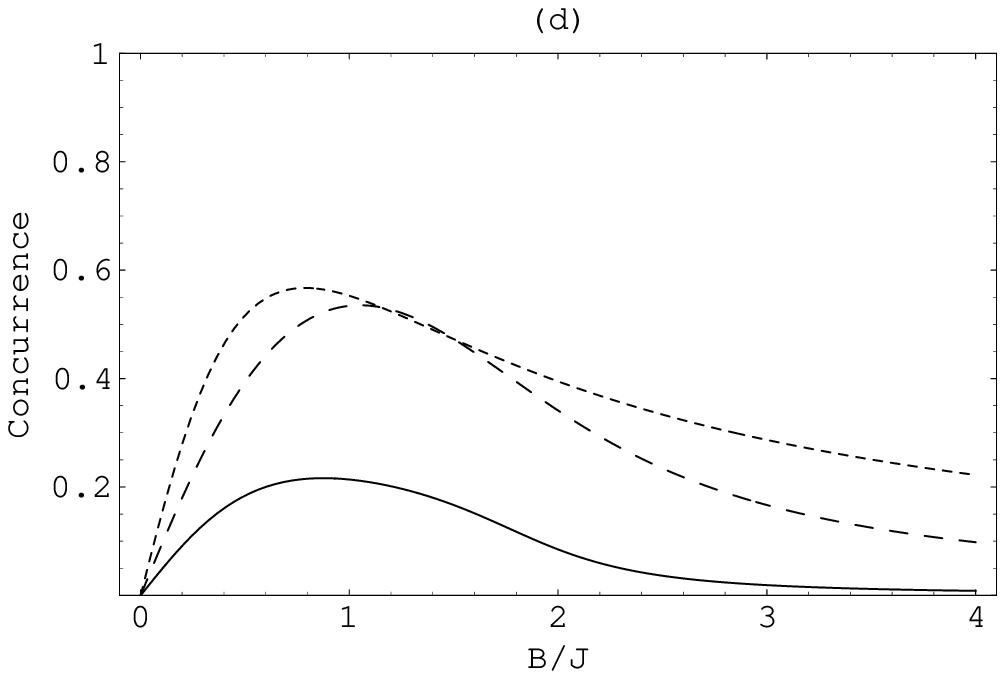,height=5cm}}}
\end{center}
\caption{Plots of concurrence as a function of the magnetic field
magnitude $B$(=$B_{1})$ in units of $J$ at zero temperature.
(\textbf{a}) Unequal field amplitudes ($B_{2}$=1.0005$B_{1})$.
Solid curve is for $\theta _1 = 0.01\pi $, long and short dashed
curves correspond to $\theta _1 = 0.1\pi $ and $\theta _1 =
0.5\pi $, respectively. Also, $\theta _2 - \theta _1 = 0.01\pi $.
(\textbf{b}) Same as (a) but with $\theta _2 - \theta _1 = 0.1\pi
$. (\textbf{c}) Same as (a) but with $B_{2}$=1.05$B_{1}$.
(\textbf{d}) The same as (c) but with $\theta _2 - \theta _1 =
0.1\pi $.} \label{fig2}
\end{figure}

\begin{figure}
\begin{center}
\centerline{\hbox{ \psfig{figure=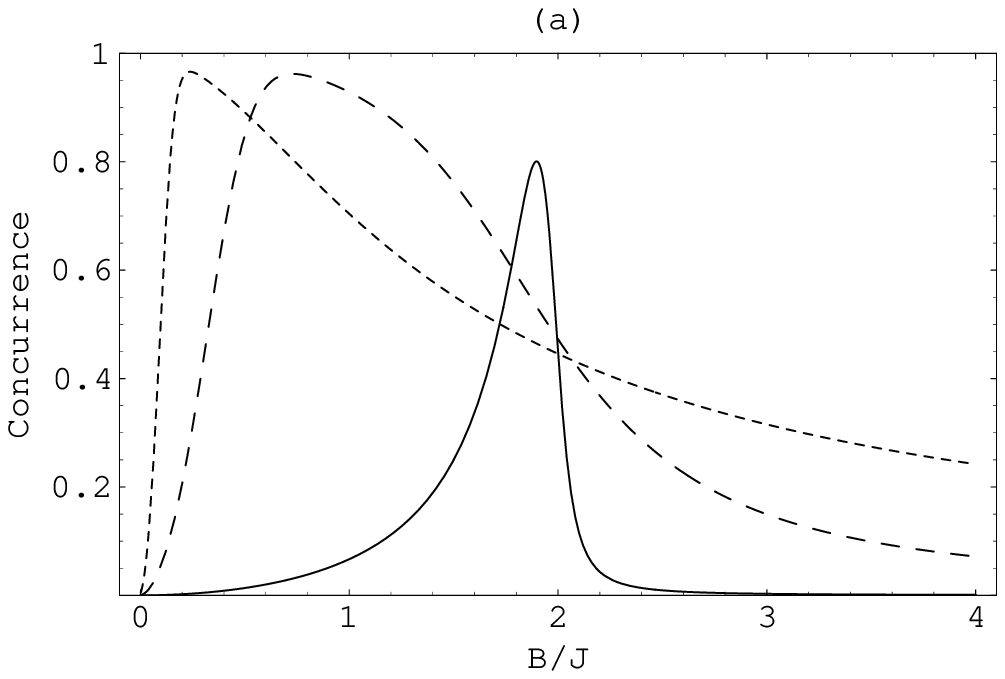,height=5cm}}}
\centerline{\hbox{ \psfig{figure=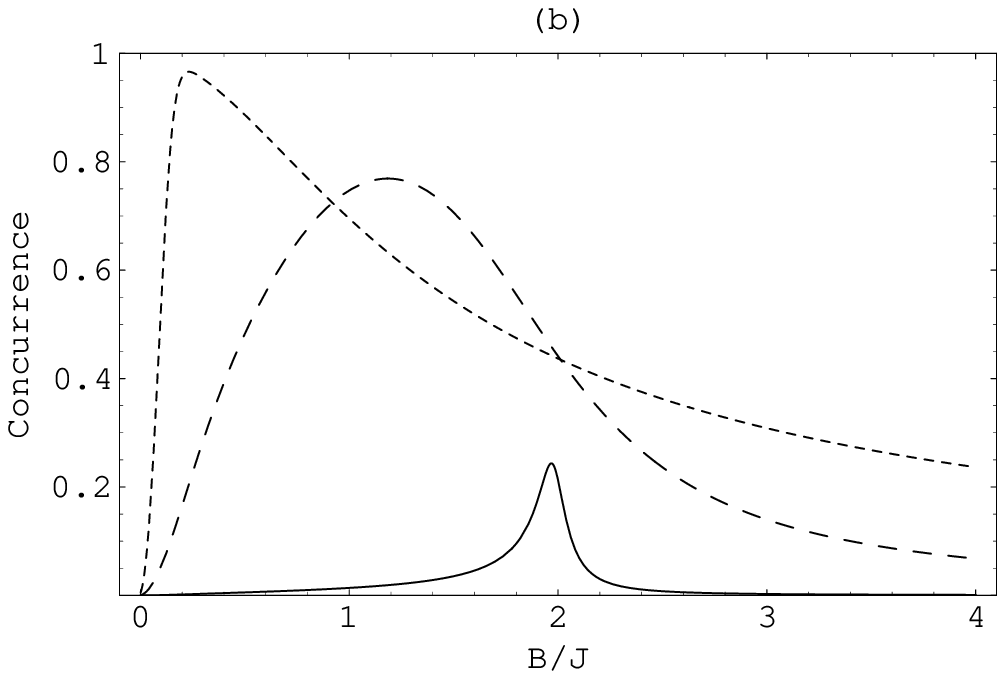,height=5cm}}}
\end{center}
\caption{Plots of concurrence as a function of the magnetic field
magnitude $B$(=$B_{1})$ in units of $J$ at non-zero temperature
($T $= 0.01$J$/$k_{B})$. (\textbf{a}) Equal magnetic field
magnitudes ($B_{1}=B_{2}=B)$. Solid curve is for $\theta _1 =
\theta _2 = 0.01\pi $, long and short dashed curves correspond to
$\theta _1 = \theta _2 = 0.1\pi $ and $\theta _1 = \theta _2 =
0.5\pi $, respectively. (\textbf{b}) Same as (a) but with unequal
field amplitudes ($B_{2}$=1.05$B_{1})$.} \label{fig3}
\end{figure}

\begin{figure}
\begin{center}
\centerline{\hbox{ \psfig{figure=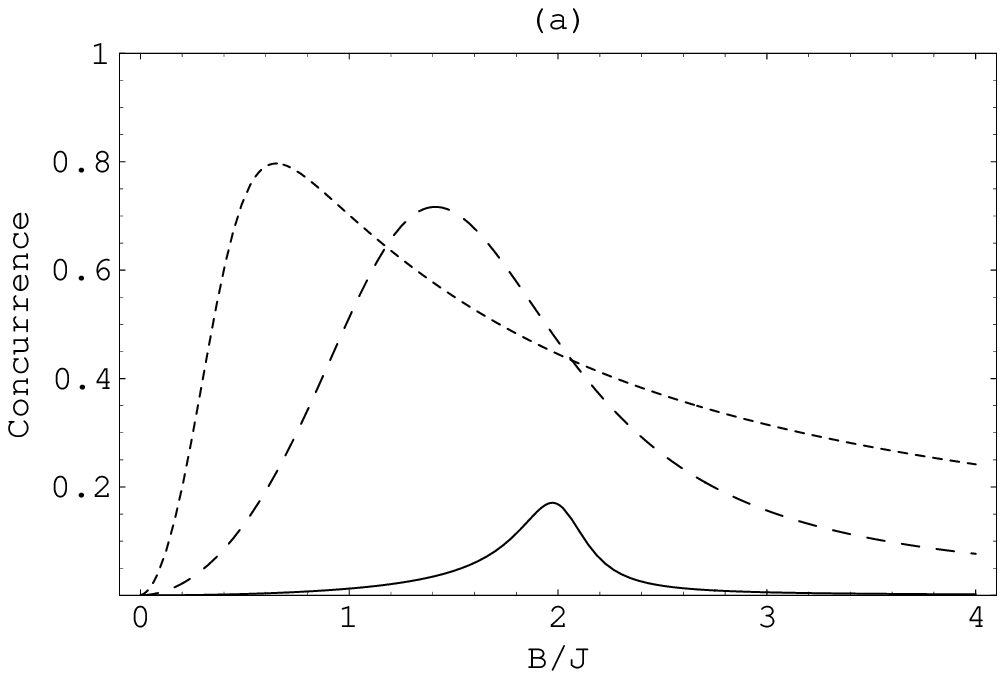,height=5cm}
\psfig{figure=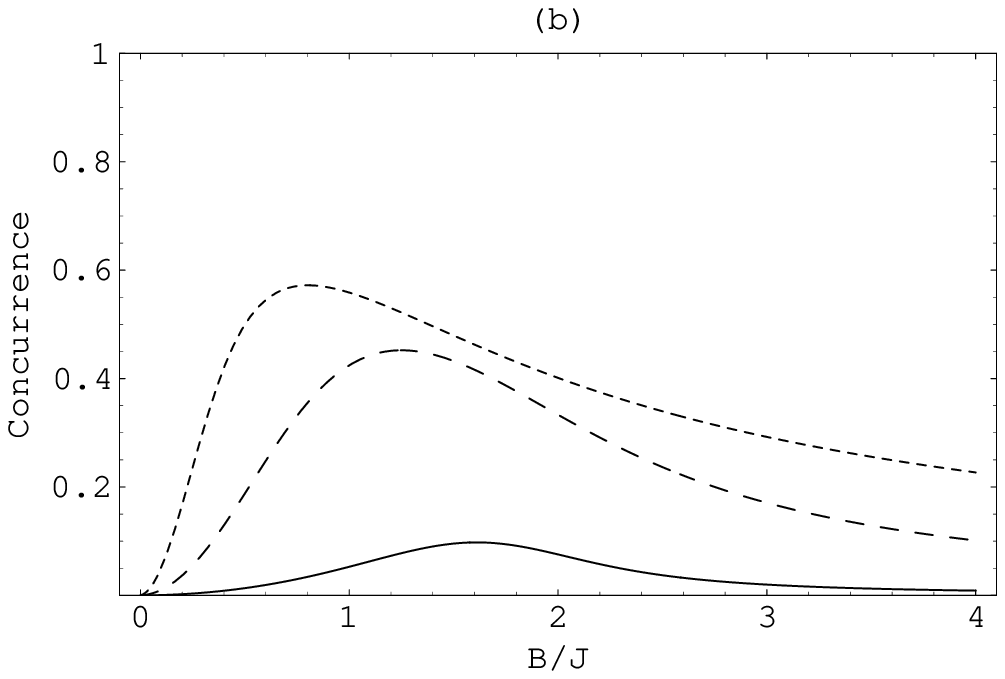,height=5cm}}} \centerline{\hbox{
\psfig{figure=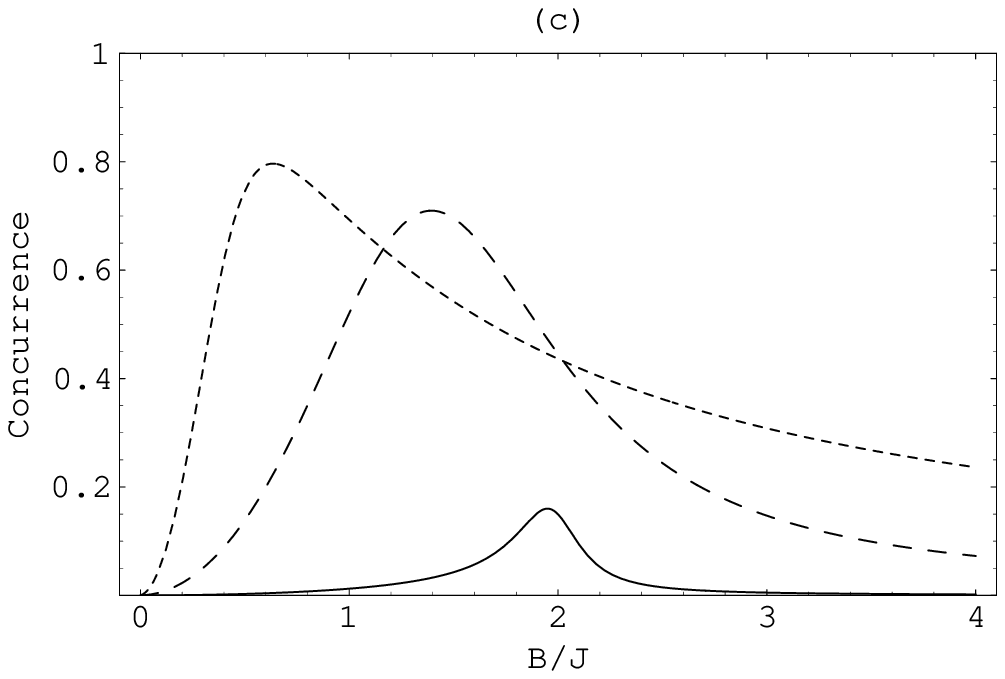,height=5cm}
\psfig{figure=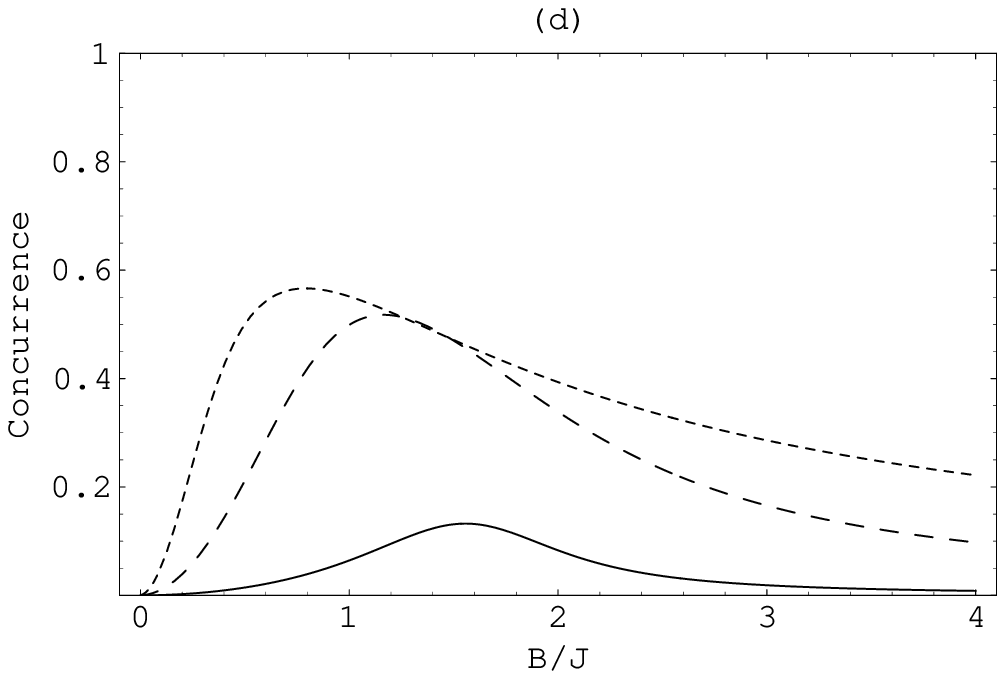,height=5cm}}}
\end{center}
\caption{The same as in Fig. 2 but with non-zero temperature $T $=
0.1$J$/$k_{B}$. } \label{fig4}
\end{figure}

\begin{figure}
\begin{center}
\centerline{\hbox{ \psfig{figure=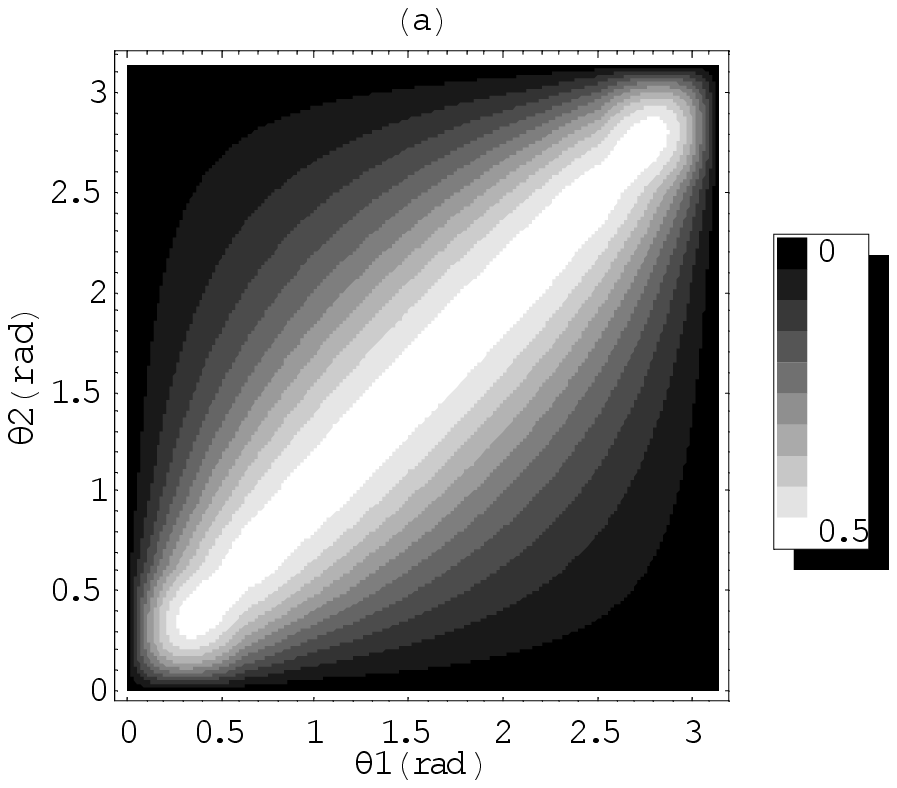,height=17.5cm}
\psfig{figure=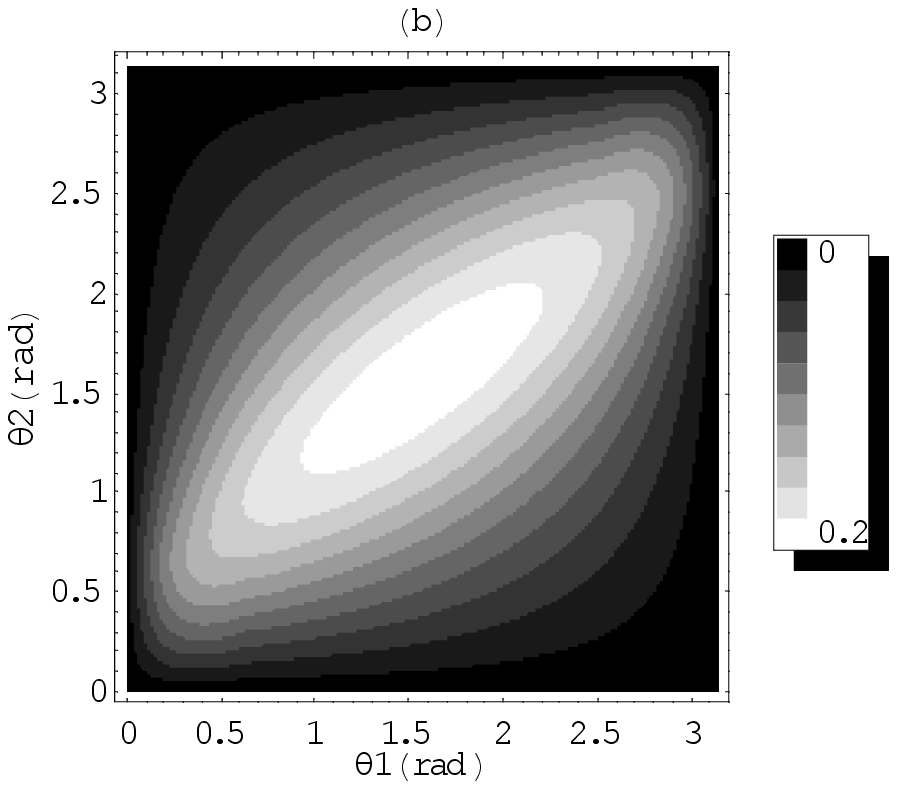,height=17.5cm}}}
\end{center}
\caption{Contour plots of concurrence as a function of the
directions of magnetic fields $\theta _1 $ and $\theta _2 $
(polar angles between field and Ising $z$-axis) for zero
temperature. In both plots $B_{1}$=2.1$J$. In \textbf{(a)}
$B_{2}=B_{1}$ and in \textbf{(b)} $B_{2}$=3$B_{1}$. } \label{fig5}
\end{figure}

\begin{figure}
\begin{center}
\centerline{\hbox{ \psfig{figure=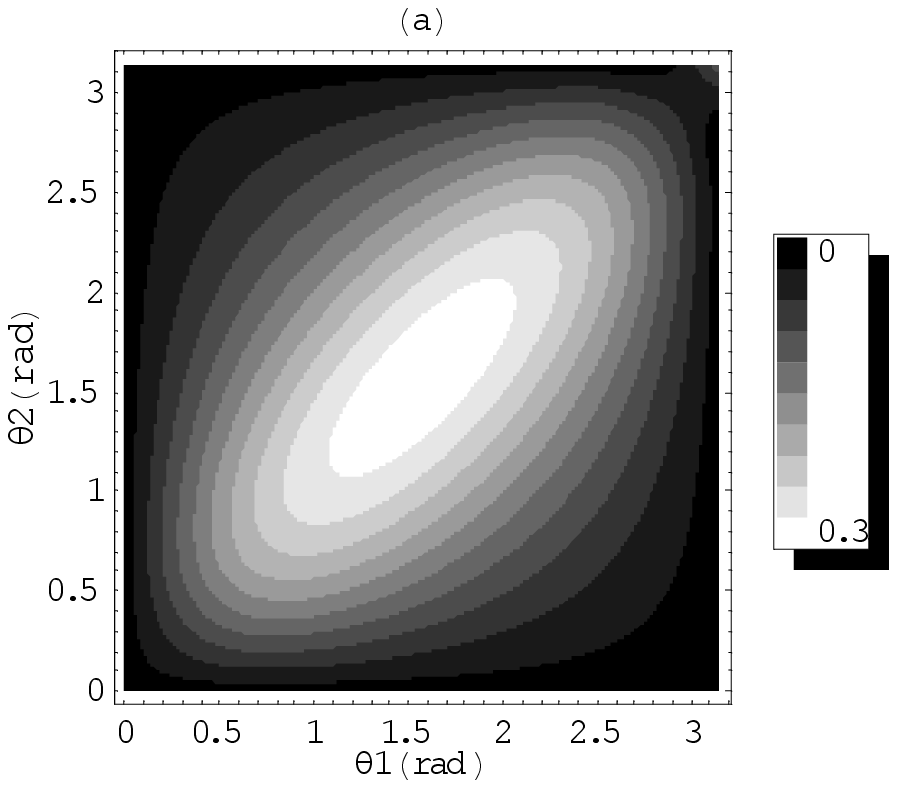,height=17.5cm}
\psfig{figure=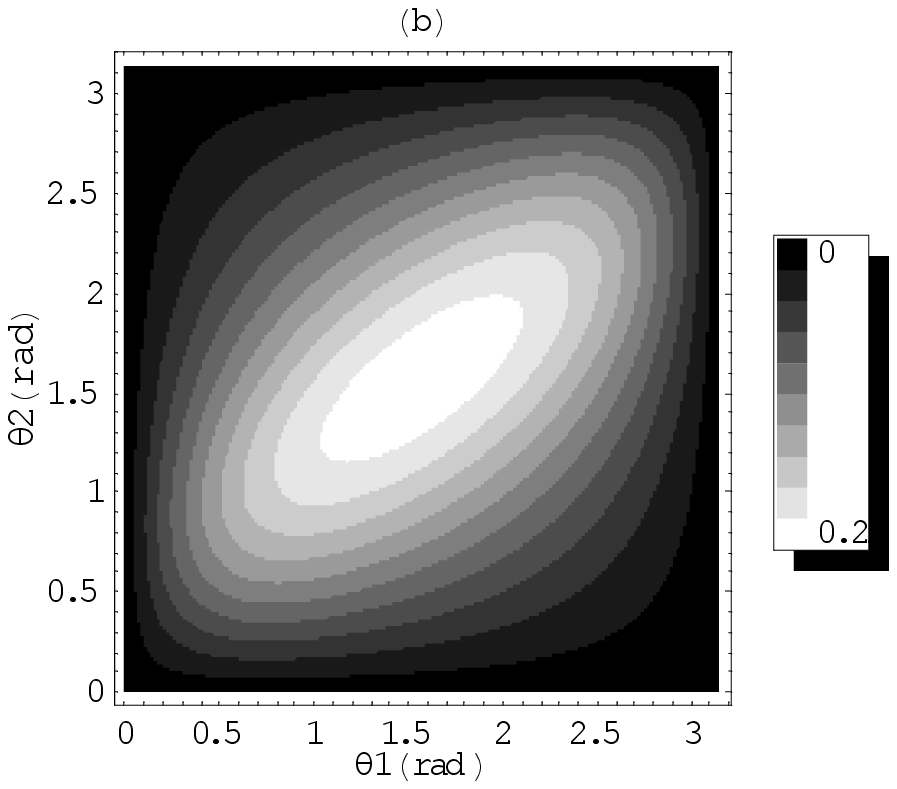,height=17.5cm}}}
\end{center}
\caption{Contour plots of concurrence as a function of the
directions of magnetic fields $\theta _1 $ and $\theta _2 $. In
both plots $T $= 1$J$/$k_{B}$ and $B_{1}$=2.1$J$. In \textbf{(a)}
$B_{2}=B_{1}$ and in \textbf{(b)} $B_{2}$=3$B_{1}$.} \label{fig6}
\end{figure}
\end{document}